\documentclass[aps,twocolumn,showpacs,showkeywords]{revtex4}

\usepackage{graphicx}
\begin{document}

\title{The nitrogen-vacancy center in diamond re-visited}
\author{N. B. Manson, J. P. Harrison and M.J. Sellars}
\address{Laser Physics Center,
Research School of Physical Sciences and Engineeering\\
Australian National University, Canberra, A. C. T., 0200, Australia}
\date{\today}

\begin{abstract}
Symmetry considerations are used in presenting a model of the electronic structure and the associated
dynamics of the nitrogen-vacancy center in diamond. The model accounts for the occurrence of
optically induced spin polarization, for the change of emission level with spin polarization and for
new measurements of transient emission. The rate constants given are in variance to those reported
previously.
\end{abstract}


\maketitle

\section{Introduction}

Since the nitrogen-vacancy (N-V) center in diamond has been detected at a single site level
\cite{Gru97,Dra99,Dra99a} the center has attracted attention for various quantum information
processing applications. For example, the center has been used as a single photon source for quantum
cryptography \cite{Kur00,Bro00} and work in this area has included impressive demonstrations
\cite{Bev02}. Another area of interest results from the center having a non-zero spin ground state.
The ground state spin can be the qubit and the optical transitions utilized for readout and for qubit
manipulation in quantum computing applications. There are again impressive demonstrations in this
area \cite{Wra01,Jel02,Jel04,Jel04a,Jel04b}.  With these successes and additional programs under
development it might be expected that the properties of the N-V center are well understood. However,
this is not the case. Despite the extensive publications the center is not well understood and the
literature contains many inaccuracies. The purpose of this paper is to re-visit our knowledge of the
nitrogen-vacancy center, provide an account of the electronic energy levels and explain the dynamics
of the center under optical excitation.

\section{Nitrogen-vacancy center}

The N-V center occurs in diamond containing single substitutional nitrogen when irradiated and
annealed \cite{Dup65,Dav76}. Electron irradiation with energies greater than 200 keV creates
vacancies \cite{Ste99}. The vacancies are mobile at 800$^{\circ}$C and can become trapped adjacent to
the nitrogen impurities. The nitrogen-vacancy complex formed has a strong optical transition with a
zero-phonon line at 1.945 eV (637nm) accompanied by a vibronic band at higher energy in absorption
and lower energy in emission. The zero-phonon line has been studied by Davies and Hamer \cite{Dav76}.
They have studied the effect of uniaxial stress and from the splitting and polarization they
established that the transition is associated with an orbital A - E transition at a site of trigonal
symmetry. The trigonal symmetry is consistent with an adjacent nitrogen-vacancy pair with $C_{3v}$
symmetry. In other studies using optical excitation Loubser and van Wyk detected electron spin
resonance (ESR) signals of a spin polarized triplet (S = 1) \cite{Lou77}. The ESR they observed was
associated with a center having trigonal symmetry and the magnitude of the optically induced signal
was found to vary with wavelength in correspondence with the strength of the A - E optical
transition. Hence, the center was attributed to the same nitrogen-vacancy complex. With an integer
spin (S = 1) the center must have an even number of electrons and it is taken that the neutral
nitrogen-vacancy complex with five electrons has acquired an additional electron from elsewhere in
the lattice, probably from another substitutional nitrogen atom. There will then be six electrons
occupying the dangling bonds of the vacancy complex \cite{Lou77}. Loubser and van Wyk proposed that
the spin polarization arises from a singlet electronic system with inter-system crossing to a spin
level of a meta-stable triplet. However, it was established from hole burning \cite{Red87}, optically
detected magnetic resonance \cite{Oor88}, ESR \cite{Red91} and Raman heterodyne measurements
\cite{Man92} that the triplet is the ground state rather than a meta-stable state. Therefore, their
model has to be modified to give a $^{3}$A ground state and a $^{3}A - ^{3}E$ optical transition. The
optically induced spin polarization can still arise from inter-system crossing and an account is
given in this paper.

The six electrons occupy the dangling bonds associated with the vacancy complex. A discussion of this
situation is included in a treatment by Lenef et.al. \cite{Len96}. Although they did not adopt the
six electron model \cite{Gos97} they did give a very useful general treatment including the six
electron situation and their presentation allows the present discussion to be brief and more
descriptive. The dangling bonds are formed from $sp^{3}$ orbitals of the carbon and nitrogen atoms
and in a vacancy approximating $T_{d}$ symmetry these can be combined to form $a_{1}$ and $t_{2}$
molecular orbitals with $A_{1}$ and T$_{2}$ symmetry, respectively \cite{Cou57}. From symmetry and
charge overlap considerations the $a_{1}$ is considered to be lower in energy and the $t_{2}$ higher.
With six electrons $a_{1}^{2} t_{2}^{4}$ will be the lowest energy configuration and this can also be
described as a $t_{2}^{2}$ hole system. When one of the neighboring carbons is replaced by a nitrogen
the T$_{d}$ symmetry will be lowered to trigonal and the $t_{2}$ orbital will be split to give, in
$C_{3v}$ notation, an $a_{1}$ and e orbital. The e hole is lower in energy and, hence, the lowest
energy configuration will be $e^{2}$, next lowest $ea_{1}$ and the $a_{1}^{2}$ highest. The
spin-orbit wave functions for the $e^{2}$ configuration give $^{3}A_{2}$, $^{1}A_{1}$ and $^{1}E$
states and the $ea_{1}$ configuration $^{3}E$ and $^{1}E$ states. The $a_{1}^{2}$ gives an
$^{1}$A$_{1}$. The optical transition is associated with triplets and so the ground state is
attributed to the $^{3}A_{2}(e^{2})$ state and the excited state to the $^{3}$E($ea_{1}$) state.
There are singlets $^{1}A_{1}$(e$^{2}$) and $^{1}E$(e$^{2}$) and $^{1}E$(a$_{1}$e) which could lie in
the same energy range as the triplets. The $^{1}$A$_{1}$(a$_{1}^{2}$) lies higher. It has been
assumed that the $^{1}A_{1}$(e$^{2}$) lies between the triplets and the present treatment accepts
this assertion and will be shown to be consistent with observation. The possibility of intermediate
$^{1}$E level(s) will be discussed.

The energy, V, of each state $^{3}A_{2}$, $^{3}$E, $^{1}A_{1}$ is determined by the above bonding
considerations and the Hamiltonian including spin-orbit, V$_{SO}$, and spin-spin, V$_{SS}$,
interaction is given by:

\begin{equation}
H = V + V_{SO} + V_{SS}
\end{equation}

Spin-orbit and spin-spin do not affect the degeneracy of the singlets whereas the spin degenerate
ground state is only affected by spin-spin interaction normally written as:

\begin{equation}
V_{SS} = \rho S_{z}^{2} + \rho^{'}(S_{x}^{2} + S_{y}^{2})
\end{equation}

where $\rho$ and $\rho^{'}$ are the axial and non-axial coefficients. The interaction spits the
ground state into a singlet, $|$A$_{2}$,S$_{z}$$>$ with symmetry A$_{1}$, and doublet,
$|$A$_{2}$,S$_{x}$$>$, $|$A$_{2}$,S$_{y}$$>$ with symmetry E. The spin states $|$S$_{z}$$>$ and
($|$S$_{x}$$>$, $|$S$_{y}$$>$) are not mixed.

The $^{3}$E state are affected by both terms. V$_{SO}$ has the form:

\begin{equation}
V_{SO} = \lambda(L_{z}S_{z})+\lambda^{'}(L_{x}S_{x}+L_{y}S{_y}) \end{equation}

where $\lambda$ and $\lambda^{'}$ are the coefficients associated with the axial and non-axial
spin-orbit interactions. The axial $\lambda$(L$_{z}$S$_{z}$) spin-orbit interaction splits the
$^{3}$E spin triplet into three two-fold degenerate states; E, E', and an (A$_{1}$, A$_{2}$) pair.
Within the $^{3}$E state non-axial spin-orbit interaction is small and will be neglected at present.
As will be discussed shortly spin-orbit can give mixing between adjacent states. This can cause a
shifting of levels which can be calculated by including this state in the calculation or by including
spin-orbit interaction to second order. The form of this latter term is the same as that of spin-spin
interaction. In high symmetry (eg T$_{d}$) this has the form \cite{Gri61}:

\begin{equation}
V_{SS} = \rho[(\underline{L} . \underline{S})^{2} + 1/2(\underline{L} . \underline{S}) +L(L+1)S(S+1)]
\end{equation}

In trigonal symmetry one has to allow for the difference in the axial and non-axial terms. The
interaction modifies the separation of the above states but does not change the wave-functions and
the wave functions are the main interest here. The states and the symmetry adapted wave functions are
given in Fig.1(a). The states with S$_{z}$ spin projection are not mixed with the spin states with
S$_{x}$ and S$_{y}$ spin projection. The $^{3}$A$_{2} \leftrightarrow$ $^{3}$E transition is
orbitally allowed and, as spin projection is not changed by the electric dipole operator, the optical
transitions will be the same strength for each of the spin projections (shown as solid vertical
arrows in Fig.1(a)). There are no transitions involving a change of spin and it can be concluded that
optical cycling of the $^{3}$A$_{2}$ - $^{3}$E transition will not result in change of spin
projection and consequently can not give any spin polarization. This situation does not change when
considering vibronic interactions as spin projection is conserved.

Spin-orbit interaction mixes singlets and triplets which transform according to the same irreducible
representation. The mixing provides an avenue whereby symmetric vibration can cause a population
relaxation between the mixed singlets and triplets. The symmetry considerations, therefore, determine
the allowed inter-system crossing and these are also shown in Fig. 1(a). Where there is only a
$^{1}$A$_{1}$ singlet level the inter-system crossing is restricted to states that transform as
A$_{1}$ irreducible representations. There can be excitation of population out of the E ground state
level with S$_{x}$ or S$_{y}$ spin projection to the A$_{1}$ spin-orbit level of the $^{3}$E state.
Population in this state can decay via the singlet to the S$_{z}$ spin projection of the ground
state. Such an excitation and decay, therefore, causes a re-orientation of the ground state spin
projection and with continuous excitation population can be transferred to the ground S$_{z}$ spin
state. This is consistent with the preferred spin orientation established experimentally
\cite{Har04}. With these selection rules, assuming the optical induced spin polarization is faster
than spin-lattice relaxation, the spin polarization would be 100$\%$ but this is not what is observed
\cite{Har06}.

The extra consideration is the non-axial $\lambda^{'}$(L$_{x}$S$_{x}$ + L$_{y}$S$_{y}$) spin-orbit
interaction which we have previously neglected. This spin-orbit term causes a mixing of the two
$^{3}$E states (denoted E and E') transforming as an E irreducible representation. These states have
different S$_{z}$ and S$_{x}$,S$_{y}$ spin projections and, hence, the mixing gives rise to optical
transitions that do not conserve spin (dashed arrows in Fig.1). The transitions are observed in hole
burning spectra \cite{Har84,Red87,Red92,Man94,San06,Mar99,Niz03a} and are the transitions that limit
the degree of spin polarization. It is anticipated that the strength of the non-axial spin-orbit
interaction is small. This is concluded from consideration of the the one electron operators.
Spin-orbit interaction will be isotropic for t$_{2}$ orbits in T$_{d}$ symmetry but the axial
contribution is quenched when the orbit is split into two-fold degenerate and non-degenerate states.
In C$_{3v}$ the non-axial spin-orbit interaction between the individual two-fold degenerate states is
formally allowed but its contribution arises from higher order effects. The non-axial spin-orbit
interaction is, therefore, anticipated to be small and the non-spin conserving optical transitions
will be weak. Although weak the transitions play an important role in limiting the degree of spin
polarization.

\begin{figure}[ht]
\includegraphics[width=0.5\textwidth]{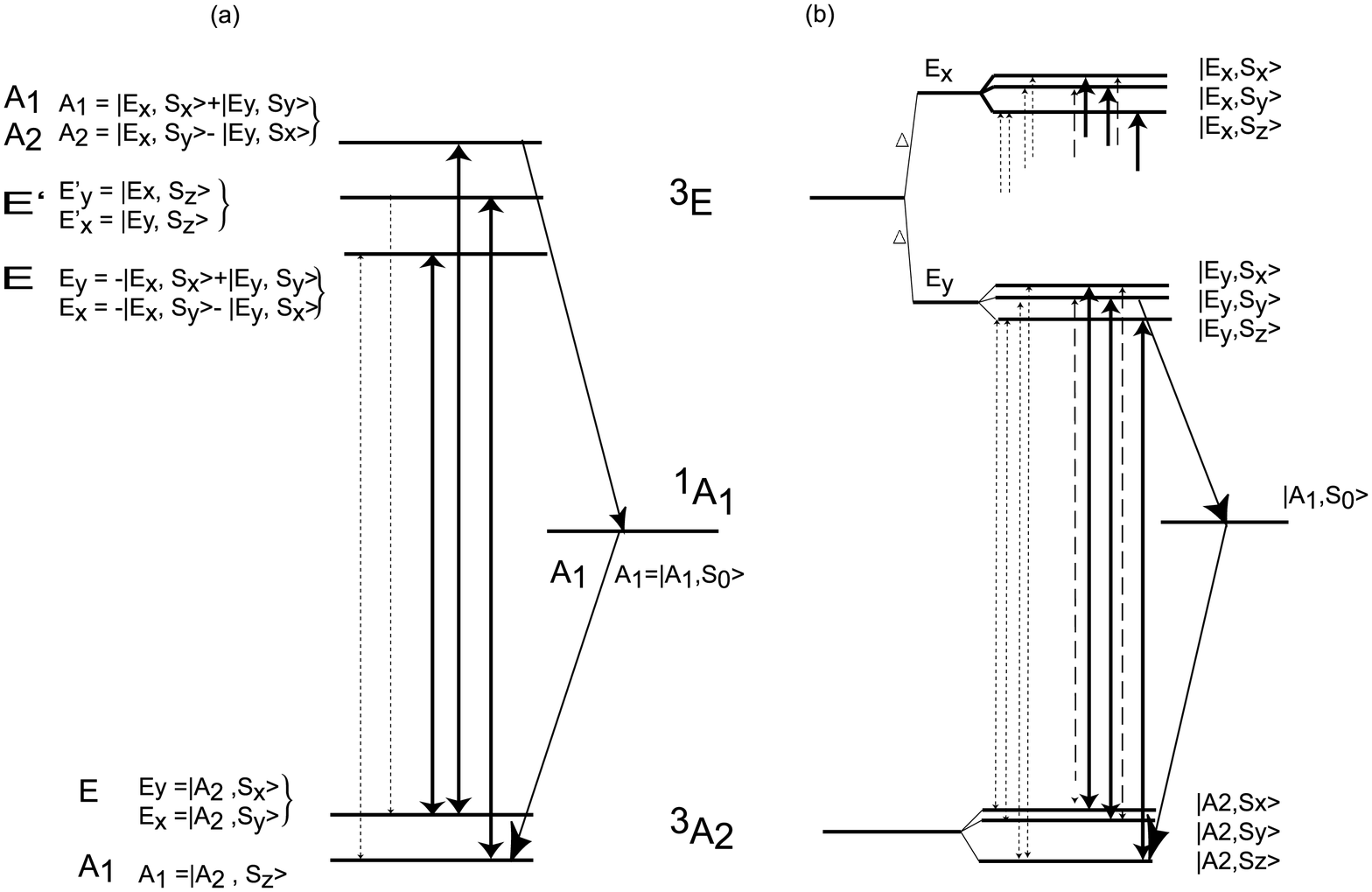}
\caption{(a) Energy levels of the N-V center in C$_{3v}$ symmetry. The excited state is split by
spin-orbit interaction whereas the ground state is split by spin-spin interaction (not shown to
scale). The figure gives the symmetry adapted wave-functions. The solid arrows indicate the
spin-allowed optical transitions. The dashed arrows indicated weak transitions which are allowed
through the mixing of the ($^{3}$E)E and ($^{3}E)$E' basis states by transverse spin-orbit
interaction. The diagonal arrows give inter-system crossing allowed by spin-orbit interaction. (b)
Energy levels and wave functions of the N-V center in the presence of a strain field. The wave
functions are appropriate for a strain field perpendicular to a reflection plane. The transitions are
derived from those allowed in (a).} \label{Fig1}
\end{figure}

We briefly consider the situation where a $^{1}$E state(s) lies between the $^{3}$A$_{2}$ and
$^{3}$E. If this were the case there could be symmetry allowed relaxation from the excited triplet
($^{3}$E)E and ($^{3}$E)E' states to such an intermediate $^{1}$E singlet state. However, using the
two hole description the relaxation is found to be forbidden in the case of the triplet ($^{3}$E)E'
state (cancellation of terms for a $^{1}$E(a$_{1}$e) and forbidden for one electron operators for
$^{1}$E(e$^{2}$) states) implying that there will still not be population transfer out of states with
S$_{z}$ spin projection. There can be relaxation from the other triplet ($^{3}$E)E state involving
the S$_{x}$, S$_{y}$ spin projections to a $^{1}$E level. Given that there is also an intermediate
$^{1}$A$_{1}$ there can be two alternate situations depending on the ordering of the $^{1}$A$_{1}$
and $^{1}$E singlet levels. Should the $^{1}$E state lie lowest the decay will be to the
($^{3}$A$_{2}$)E component of the ground state. This state has a S$_{x}$, S$_{y}$ spin projection and
the optical cycling will have had no consequence implying no change in spin orientation. However, if
the $^{1}$E state lies above the $^{1}A_{1}$ level there will be radiative (or non-radiative) decay
to the lower singlet level, the $^{1}A_{1}$, followed by the relaxation, as discussed in the previous
paragraph, to the S$_{z}$ spin projection of the ground state. This process will lead to the same
spin change and spin polarization as before. The simple consequence of involving the $^{1}$E singlet
is that the total ($^{1}A_{1}$ plus $^{1}E$) spin polarization process will be more efficient. Thus,
if there is an intermediate $^{1}$E state, to be consistent with observation it must lie at an energy
higher than the $^{1}A_{1}$ state. It is recognized that this order is in disagreement with
calculated energy levels \cite{Gos94}. However, there is an appeal of including a higher single
$^{1}E$ state as it could lie close to the excited triplet level and the $^{1}A_{1}$ close to the
ground state thus accounting for the relatively fast inter-system crossing reported below. However,
there is no fundamental difference in the dynamics and it is sufficient in this work to restrict the
discussion to one singlet level, a $^{1}$A$_{1}$.

It is common for there to be strain in diamond and it is worth considering the consequence to the
energy levels and the associated dynamics. There will be no fundamental change with axial strain as
all it gives is a uniform shift of the energy levels and no change of the wave functions or the
dynamics. However, the component of the strain at right angles to the axis of the center lowers the
symmetry and the extra crystal field lifts the orbital degeneracy of the excited state to give two
orbital non-degenerate states denoted E$_{x}$ and E$_{y}$ \cite{Len96}. We consider the case where
this strain splitting is larger than the spin-orbit interaction. Where the strain retains a
reflection plane the Z, X, Y axes will be determined by symmetry. The wave functions for this
situation are as given in Fig.1(b). The diagonal spin-orbit interaction is quenched and the order of
states (same in both optical components) are determined by spin-spin interaction. In the limit of
small non-axial spin-orbit interaction there is still little mixing of the S$_{x}$, S$_{y}$ spin
states with the S$_{z}$ spin states. The transitions and inter-system crossing can be determined from
the parent state and are shown in Fig. 1(b). The spin allowed transitions will have near the same
strength as in the zero strain case but they are now totally polarized. The only minor change is loss
to the dashed transitions between the states with different S$_{x}$ and S$_{y}$ projections. These
arise where the strain is not sufficient to totally quench the effect of the diagonal spin-orbit
interaction. A more significant effect is in the strength of the non-spin-conserving transitions
induced by non-axial spin-orbit interaction. They will become significantly stronger as the
separation of the S$_{z}$ and (S$_{x}$, S$_{y}$) is reduced and the mixing increased. When the strain
does not retain reflection symmetry the effective X and Y axis may be different between ground and
excited states and all transitions and inter-system crossings will become allowed in principle (no
symmetry restrictions). However, the selection rules will be dominated by those allowed in zero order
and the perturbation approach in Fig. 1(b) is anticipated to give a reasonable approximation to the
dominant excitation and decay channels.

Which of the diagrams, Fig.1(a) or Fig.1(b) (or an intermediate case), is appropriate for a given
center depends on the relative magnitude of the stress, spin-orbit and spin-spin interaction.
Spin-orbit for the carbon atom is known to be a few cm$^{-1}$ ($\sim$200 GHz) and a spin-orbit
splitting of 1 cm$^{-1}$ (30 GHz) was obtained for the $^{3}$E state from optical magnetic circular
dichroism measurements \cite{Red87}. This value maybe marginally high as an optical line width of 15
GHz has been reported recently for small ensembles within single crystals of diamond \cite{San06} and
there is an indication that spin-spin interaction may also contribute to the line width. Santouri et.
al. \cite{San06} have also shown, using two-laser hole burning experiments, that for even a small
strain splitting of 10 GHz the energy scheme is equivalent to Fig. 1(b). As strain splitting is
usually much larger ($>$ 100GHz) it is taken that this energy scheme will be typical of centers in an
ensemble sample as used here.

The dynamics associated with optical excitation can be determined without detailed knowledge of the
excited state energy levels. This is because from the above discussion it can be seen that the system
can be reasonably quantized by its spin projection S$_{z}$ or (S$_{x}$, S$_{y}$) and this is not
altered by stress. The dynamics are largely determined by the crossing between these two spin
projections and the symmetry considerations give the two main mechanisms. The changes of spin are
through the inter-system crossing via the singlet and through the weak non-spin-conserving optical
transitions. The effective energy scheme is given in Fig.2 where the S$_{z}$ states are shown on the
left and the (S$_{x}$, S$_{y}$) on the right. The singlets are drawn centrally. Spin polarization
involves displacement of population from the states on the right to the states on the left. The
transitions including those that change the spin projection are shown as solid lines as determined by
the parent states in C$_{3v}$ symmetry. For completeness other transitions allowed in low symmetry
are shown as dashed lines. The low symmetry situation is similar to that proposed by others
\cite{Niz03, Jel04a, Niz05} and to assist comparison we adopt their shortened notation. The ground
spin states are defined as x, y and z; the excited triplet states as x', y' and z' and the singlet
level as s. In the following experiments we determine the values of the parameters using bulk
samples.

\begin{figure}
\includegraphics[width=0.5\textwidth]{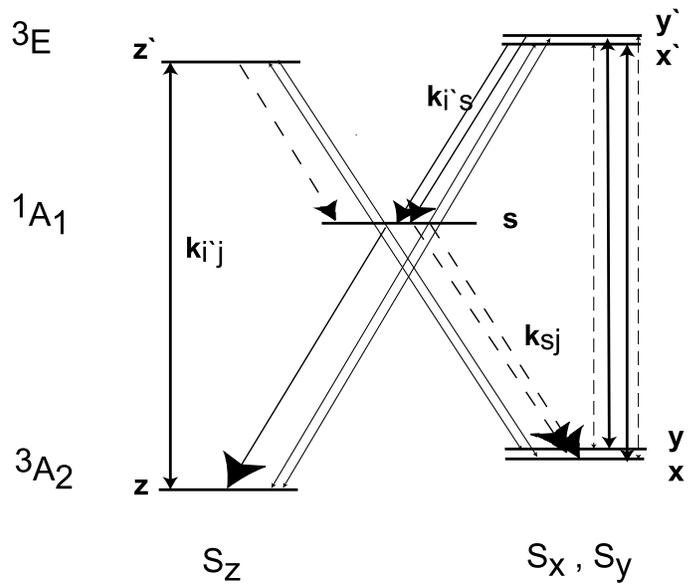}
\caption{Energy levels for a perturbed N-V center. The significant energy levels of Fig.1(b) are
re-drawn to highlight the inter-system crossing. All the radiative and non-radiative transitions are
shown. The allowed transitions considered in the text are shown by solid arrows. The dashed vertical
arrows on the right are allowed but can be taken to be zero for present experiments. Also the
inter-system crossing indicated by the dashed arrows are zero in first order.} \label{Fig2}
\end{figure}

\section{Experimental details}

Several type 1b diamonds were used in these studies. They were irradiated with energetic electrons
($>$ MeV) and annealed. The nitrogen-vacancy concentrations varied from 3 x 10$^{18}$ per cm$^{3}$ to
10$^{17}$ per cm$^{3}$ and there was no indication that the dynamics of the optical cycle varied
significantly for concentrations in this range.

The experiments involve transient and CW excitation studies using a frequency doubled Nd:YAG laser.
The excitation wavelength is 532 nm which is near the peak of the $^{3}$A$_{2}$ $\leftrightarrow$
$^{3}$E absorption band. The emission was detected by a S-20 photomultiplier or pin diode with a
response time of 30 ns. With the exception of the measurement of the spectrum the emission wavelength
detected was selected using absorptive filters.

The studies involved three different excitation and detection geometries. For low level excitation
the intensity was determined from the laser power and the beam diameter. The emission at 45$^{o}$ was
focussed on the pin diode. For high intensity measurements a confocal arrangement with an oil
immersion objective was used. The excited diameter was 0.2 micron with the emission spot focussed
onto a pinhole. A 1 mW beam gave an estimated power density of 2 x 10$^{6}$ W/cm$^{2}$. With the
small excitation and detection volumes the signals were weak and long collection times were required
to obtain satisfactory signal to noise ratios. The majority of the measurements were taken with a
third geometry where satisfactory signals could be obtained more rapidly. The light was focused with
a microscope objective and the back emission collected by the same objective. For the same laser
power the intensity at the sample was approximately two orders of magnitude lower than that obtained
with oil immersion objective and it required a 100 mW laser beam to obtain a maximum intensity of 2 x
10$^{6}$ W /cm$^{2}$. The unsatisfactory feature of this geometry is that the intensity was not
constant over the collected volume.

At low intensities where slow ($>$ms) responses were obtained the light was gated with a mechanical
chopper whereas for the fast speeds associated with the high intensities the light was gated using
two acousto-optic modulators in series. The rise time of the A-O modulators is 30 ns. When gating off
there was a weak component that lasted for a $\mu$s. However, no measurements were taken when gating
the light off.

\subsection{Preliminary measurements}
The emission of the N-V center has been reported on many occasions. The general characteristics are
shown in Fig.3. Near room temperature the emission gives a band stretching from 630 nm to 800 nm with
vibrational structure and a weak zero-phonon line at 637 nm (Fig.3(a)). Cooling has little effect on
the vibronic absorption band \cite{Dav74} and consequently there is little change in the amount of
light absorbed from the laser beam when using an excitation wavelength within the vibronic band.
Consistent with this, the total emission does not show a significant change when the temperature is
lowered (insert in Fig.3(a)). The most obvious change in cooling to low temperatures is that the
zero-phonon line becomes sharper and more prominent (Fig.3(b)). The Huang-Rys factor is 3.7 with only
2.7 percent of the transition strength being associated with the zero-phonon transition \cite{Dav74}.
At high excitation densities there can be photo-ionization of the N-V center of interest and the
creation of the neutrally charged [N-V]$^{0}$ center \cite{Man05,Gae05}. This center has a
zero-phonon line at 575 nm with a vibronic band to lower energy \cite{Dav79}. The increased
contribution of this center at high intensities is illustrated in Fig.3(b).

\begin{figure}
\includegraphics[width=0.5\textwidth]{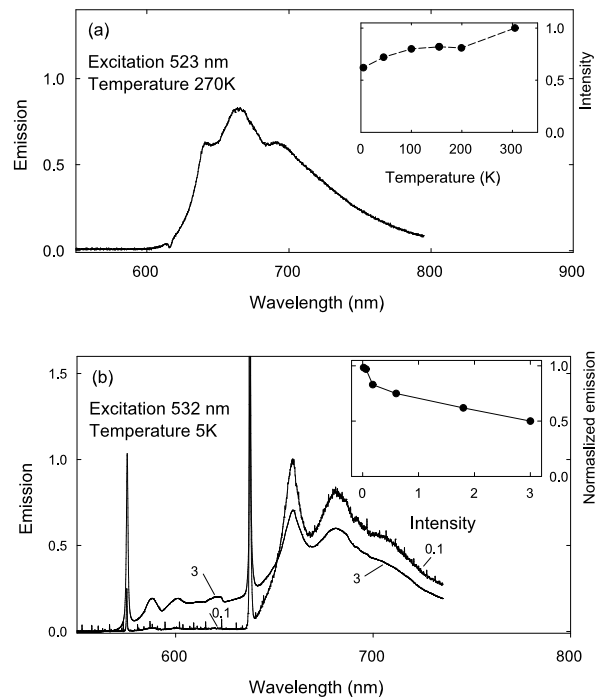}
\caption{Spectral characteristics of the N-V Center. (a) Emission spectra of a diamond containing a
high concentration of N-V centers measured near room temperature using low excitation densities of 1
W /cm$^{3}$. Insert: the variation of the N-V emission at 666 nm as a function of sample temperature
(b) Low temperature emission spectra of the N-V center measured with various excitation densities
(from reference \cite{Man05}). Insert: variation of N-V emission intensity as a function of
excitation energy density (units of 1 x 10$^{5}$ W /cm$^{3}$). Emission normalized to excitation
intensity.} \label{Fig3}
\end{figure}

As discussed earlier the N-V center exhibits optically induced spin polarization of the ground state
triplet and there is a change of the emission intensity associated with this polarization (no change
in absorption). The change is illustrated in Fig. 4.  In the dark the sample becomes unpolarized and
when excited the emission has an initial intensity level A. After exciting for a period the sample
becomes polarized and the strength of the emission increases to a second value B (Fig.4(a)). The rate
at which the emission increases from the level A to B is linearly dependent on the excitation
intensity. Should the sample be in the dark for a period less that the spin-lattice relaxation time
(T$_{1}$) the initial emission level will be at a value between A and B and varying the dark period
varies this level. This variation of emission level with the duration in the dark can be used to
establish the spin lattice relaxation time and measurements of this type are shown in Fig.4(b). The
A/B ratio is of considerable interest. To ensure a satisfactory A measurement it is necessary to have
the sample in the dark for a period long compared to T$_{1}$ whereas to obtain the saturation value B
the intensity has to be sufficient to obtain the higher emission level within a time short compared
to T$_{1}$. A decrease of the spin polarization through spin diffusion must also be avoided. When
meeting these conditions the A/B ratio is 0.86 $\pm$0.02.

\begin{figure}
\includegraphics[width=0.5\textwidth]{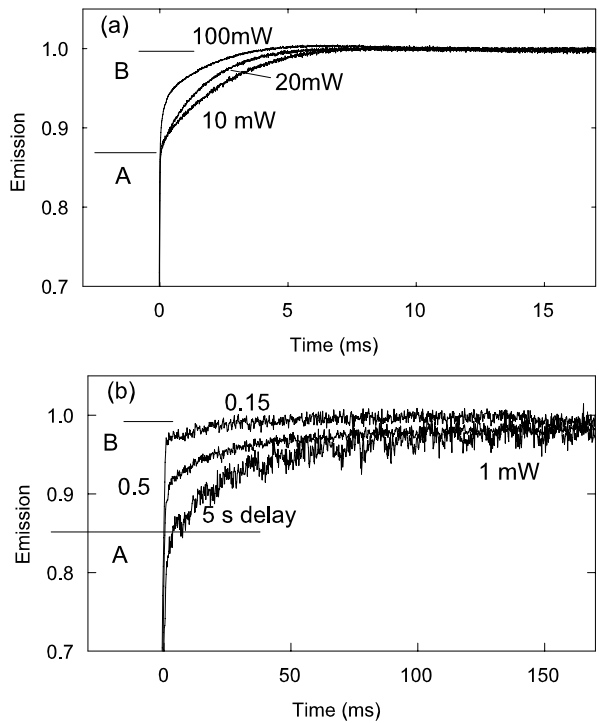}
\caption{Emission obtained when gating on excitation at time 0. The excitation is at 532 nm and is
gated with a mechanical shutter. (a) Response for various laser powers; 10 mW corresponds to an
intensity of 1 W/cm$^{2}$. The emission increases to an initial value A, 86\% of its final value. The
rate of increase of the emission to its final CW value B depends on intensity of excitation. Sample
has spin-lattice relaxation time of 100 ms and sample held in dark for 500 ms prior to excitation.
(b) Responses for various periods in the dark. For these traces the sample is cooled and the
spin-lattice relaxation time is increased to 500 ms.} \label{Fig4}
\end{figure}

\subsection{Dual pulse measurements}

The sample is excited with two intense excitation pulses and the delay between two pulses is varied.
These measurements utilized the confocal arrangement with oil immersion and intensities of 10$^{6}$
W/cm$^{2}$. The repetition rate was 10 kHz. The results are shown in Fig.5(a). In the figure the
response associated with the first pulse of every pair is overlapped whereas the emission associated
with the second pulse is displaced as the delay between the pulses is varied. Every pulse exhibits an
initial peak followed by a decay within a $\mu$s to a lower level. For the first pulse the magnitude
of the peak is 1/3 that of the equilibrium signal. When the light is gated off the sample has to
remain in the dark for a period before the second pulse gives a peak and the magnitude of the peak
increases as the duration of the dark period is increased. This recovery has two components and the
faster recovery is shown to have a response time of 0.3 $\mu$s (Fig.5(b)). There is a slower recovery
over the 100$\mu$s between pulses and we will comment on this slower recovery later.

\begin{figure}
\includegraphics[width=0.5\textwidth]{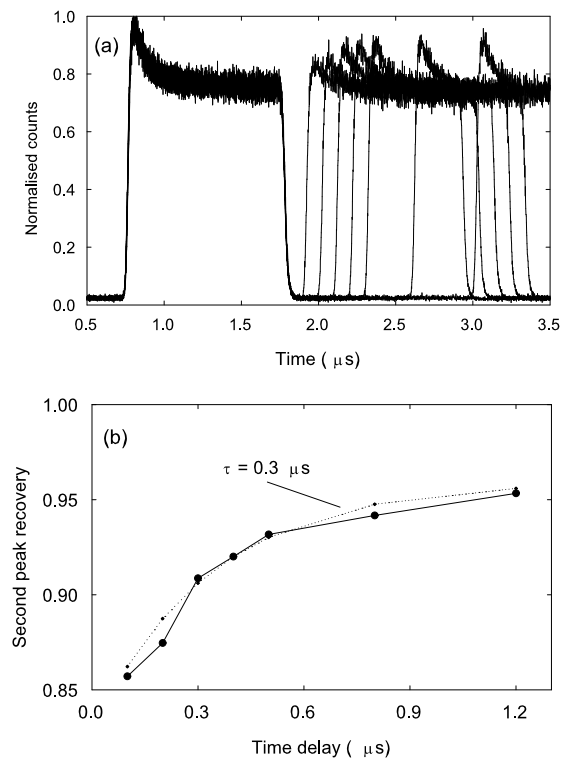}
\caption{Emission response to a double light pulse (a) The emission is shown with various delays
between the two pulses. The responses for the first pulses all overlap whereas the second pulse is
delayed by the dark period. The peak associated with the second pulse recovers with the dark period
and the height of this pulse is plotted in (b) as a function of the duration in the dark. The dashed
curve illustrates the response for a 0.3 $\mu$s recovery rate. The repetition rate of the pulse pair
was 10 kHz. The intensity of the excitation was 3 x 10$^{6}$ W /cm$^{2}$.} \label{Fig5}
\end{figure}

For the second series of measurements the emission is obtained for a pulse pair of 1$\mu$s duration
separated by 1$\mu$s but with long delays between the pulse pairs. The measurements were made using
the alternate geometry where the excitation and detection involved a larger but less well defined
volume. The period between pairs ($>$10 ms) was chosen to be much larger than ground state
spin-lattice relaxation time, T$_{1}$ (1 ms). The intention is for the system to be unpolarized at
the start of the first pulse but polarized at the start of the second pulse. Fig. 6(a) shows the
results of the emission response for pulse pairs for laser powers from 5 mW to 200 mW corresponding
to estimated intensities of 10$^{5}$ W/cm$^{2}$ to 4 x 10$^{6}$ W/cm$^{2}$. The emission is
restricted to longer wavelengths ($>$ 700 nm) to avoid including emission from [N-V]$^{0}$ centers.
The emission of the [N-V]$^{0}$ center by itself was also obtained by detecting the emission at 590
nm using a narrow band filter. This emission is shown in Fig.6(b).

\begin{figure}
\includegraphics[width=0.5\textwidth]{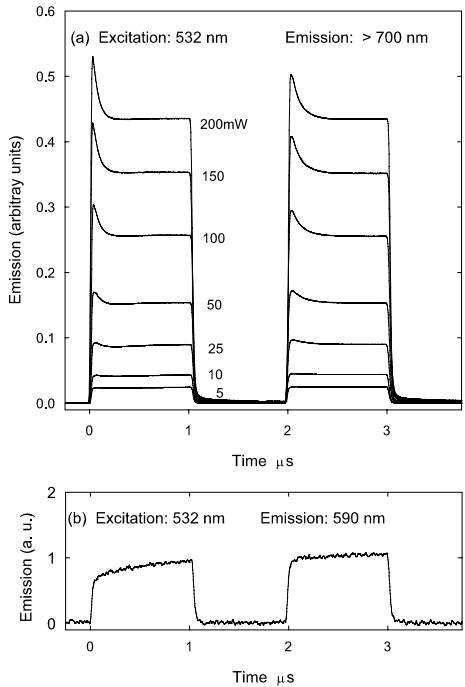}
\caption{Emission of the N-V center for a pair of square wave excitation pulse of light at 532 nm.
The light is focused with a microscope objective and for 100 mW the intensity is 2 x 10$^{6}$
W/cm$^{2}$ The back scattered emission is detected using a (a) 700 nm long pass filter, (b) 100 nm
band pass filter at 590 nm.} \label{Fig6}
\end{figure}

The responses in Fig.6(a) show peaks at the start of each pulse and the magnitude increases with
intensity. The peak height of the second pulse is consistently several percent lower than that of the
first. Also the decay rates are different for the two pulses, the first being faster than the second.

The effect of applying a weak magnetic field of a few hundred gauss was also recorded. The response
to a pair of intense (100 mW) excitation pulses at 532 nm were measured both with and without the
magnetic field  applied in a random direction (Fig.7). In comparing the two traces there are three
significant differences. The first emission peak is almost the same for the two traces but the
subsequent decay is to a much lower level when the magnetic field is applied. With the second pulse
there is a significant difference in the peak heights, being lower with the field is applied. Also
with the second pulse the decay to the lower level is faster when the magnetic field is applied.

\begin{figure}
\includegraphics[width=0.5\textwidth]{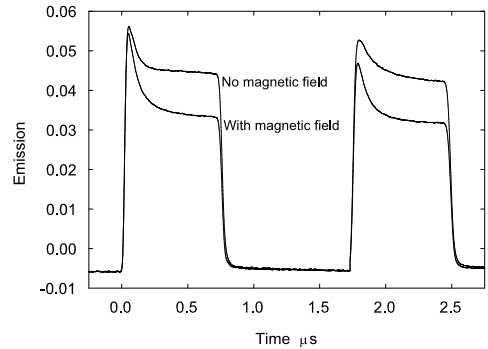} \caption{Emission of the N-V center for a
pair of square wave excitation pulses of light at 532 nm with laser intensity of 100 mW. No magnetic
field is applied in the case of the upper trace and corresponds to the 100 mW trace in Fig. 6. For
the lower trace a field of 500 gauss is applied in a random direction (not aligned with an axis of
any center).} \label{Fig7}
\end{figure}

\section{Relaxation and Inter-system Decay Rates}

The dynamics scale with the lifetime of the excited state and various values have been reported in
the literature. Collins et al \cite{Col83} have obtained values of 12.9 $\pm$ 0.1 ns for a natural
diamond and 11.6 $\pm$ 0.1 ns for a synthetic diamond. Lenef et.al. \cite{Len96} has also obtained a
measure of 12.96 $\pm$ 0.14 ns obtained in relation to photon echo measurements.  A value of $\tau$ =
13 ns is adopted here. The radiative rate constants are, hence, k$_{x'x}$ = k$_{y'y}$ = k$_{z'z}$ =
77 x 10$^{6}$ s$^{-1}$. (The present experiments are not sensitive to the dashed vertical transitions
in Fig.2 between the x, x' and the y, y' states and the associated rate constants can be taken to be
zero, ie k$_{xy'}$ = k$_{yx'}$ = 0). It is noted that the present model indicates that there should
be two components to the emission decay associated with excited states, one with z' and one with
x',y'. The emission from z' excited state has the slower decay rate and dominates when the system is
spin polarized, perhaps explaining why the two values (differing by only 30$\%$ ) have not been
detected. Two components of 9 ns and 2 ns have been observed by Hanzana et al \cite{Han97} but the
measurements are inconsistent with previous values. These lifetimes require further investigation.

The increase in emission from a level A to a level B can be used to establish the fraction of the
population transferring into the singlet. For example the 14\% emission change between having all the
population in the z spin ground state and having the population evenly distributed between the three
spin states implies 27 \% of the population from each of the x', y' excited spin states transferring
non-radiatively to the singlet, s. This has to be increased to 39\% to allow for the incomplete spin
polarization reported below. The inter-system crossing rates are then k$_{x's}$ = k$_{y's}$ = 30 x
10$^{6}$ s$^{-1}$. In correspondence with the model, the inter-system crossing from the z' state to
the singlet is taken to be zero and, hence, k$_{z's}$ = 0.

For the low intensities no population is maintained in the singlet. This is changed at high
intensities and the transient emission displays different characteristics. With population being
stored in the singlet level there is a drop in emission and this is observed in all of the two-pulse
experiments (Figs.5, 6 and 7). There will be no initial peak if the excitation is gated on and off
within a few ns as there will be no change in the singlet population. The recovery of the peak
requires a period in the dark (Fig.5) and the time required corresponds to the rate at which
population decays from the singlet to the ground state. The value of the singlet lifetime obtained
from this peak recovery is 0.3 $\mu$s and this gives k$_{sz}$ = 3.3 x 10$^{6}$ s$^{-1}$. As in the
model we take k$_{sx}$ = k$_{sy}$ = 0.

In a recent paper we have reported that the maximum spin polarization obtained for an ensemble under
continuous excitation is 80$\%$ \cite{Har06}. This means that the probability of optically
transferring spin projection from the z spin state to an x, y spin is 1/4  of the above process
giving rise to the spin polarization. The mechanism is attributed to the non-spin conserving optical
transitions (diagonal arrows between triplet levels in Fig. 2) and implies that the rate constants
are k$_{zx'}$ = k$_{zy'}$ = k$_{x'z}$ = k$_{y'z}$ = 1.5 x 10$^{6}$ s$^{-1}$. Loss of spin
polarization could also arise from the reverse inter-system crossing via the singlet level. However,
the model predicts that the rate is zero, k$_{z's}$ = 0 and, as given previously, the decay from the
singlet to the x and y ground states are also zero.

\section{Rate equations}

In the previous section estimates of the parameters of the model in Fig.2 have been obtained from
simple experimental observations. By adopting these parameters we can determine the populations and
the emission for any optical field by solving the classical rate equations:

\begin{equation}
 dn_{i}/dt = \Sigma_{j}(k_{ji}n_{j} - k_{ij} n_{i})
 \end{equation}

where n$_{i}$ is the population of level i  (i = z, x, y, z', x', y', s). and k$_{ij}$ gives the rate
for the i $\rightarrow$ j transition. The significant parameters associated with the optical
transitions are:

\begin{equation}
(k_{z'z}, k_{x'x}, k_{y'y}), (k_{x'z}, k_{y'z}, k_{z'x}, k_{z'zy}), (k_{x'y}, k_{y'x})
\end{equation}

 where the values within brackets are equal in first order. The related
optically driven terms are obtained by setting k$_{ij}$ = k * k$_{ji}$ where k indicates the strength
of the optical pumping. k = 1 corresponds to the case where the optical pumping rate of the allowed
transitions equals the emission decay rates. The inter-system crossings are determined by the rates:

\begin{equation}
(k_{z's}),  (k_{x's}, k_{y's}), (k_{sz}),  (k_{sx}, k_{sy})
\end{equation}

There are no reverse terms associated with the inter-system crossing and, hence, the related
parameters with the indices reversed are zero. Likewise all the relaxation rates between the spin
levels z, x and y and between levels z', x' and y' are small. These parameters can be considered
equal and given a small value but the effects are not significant in the calculated responses.

Where a sample has been in the dark for a period long compared to the spin-lattice relaxation time
the population will initially be equally distributed over the three ground state spin levels, z, x
and y. Emission is established in a time of the order of the excited state lifetime of 10 - 20 ns and
with continuing excitation the emission level increases as population is transferred to the z state
(Fig.8). This behavior is in correspondence with observation (Fig.4). However, little significance
can be drawn as there has not been an independent measurement of the optical pumping rate and the
magnitude of the rise in Fig.4 has been used in determining the parameters of the system.

\begin{figure}
\includegraphics[width=0.5\textwidth]{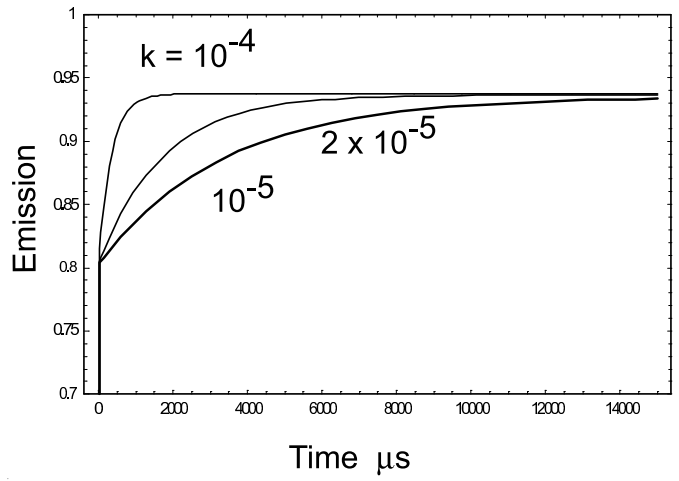}
\caption{Emission calculated from solution of the rate equation for energy scheme in Fig. 2. The
intensities k are given in units of 1/$\tau$. The value of the parameters in units of 10$^{6}$
s$^{-1}$ introduced in the text are k$_{zz'}$ = k$_{xx'}$ = k$_{yy'}$ = 77; k$_{sz}$ = 3.3, k$_{sx}$
= k$_{sy}$ = 0; k$_{z's}$ = 0; k$_{x's}$ = k$_{y's}$ = 30; k$_{zx'}$ = k$_{zy'}$ = k$_{xz'}$ =
k$_{yz'}$ = 1.5} \label{Fig8}
\end{figure}

Other than the optical pumping rate there are no free parameters when calculating the emission
associated with the dual pulses experiments. In these experiments the intensities are high and the
transfer of population into the singlet level is initially faster than the relaxation from the
singlet to the ground state. Consequently there is a build up of singlet population and associated
with this there is a drop in the emission level. The situation is calculated for a light field
switched on and held constant for 1$\mu$s, switched off for 1$\mu$ and then switched on again for a
further 1$\mu$s. With the intense excitation the system reaches equilibrium during the 1$\mu$s pulse
and so is spin polarized well before the end of the first pulse. In the dark period there is a
relaxation of the singlet population but there is no loss in spin-polarization. The result is that
for the second light pulse the system starts spin polarized with a preferential population in the z
spin ground state. In this case the transfer to the singlet is less efficient and the build up of
population in the singlet level is slower. This accounts for the observed slower drop in emission
intensity with the second pulse. The behavior for representative intensities is shown in Fig. 9.

The results of the calculations can be compared with the experimental measurements of Fig. 6. It
should be recognized that the calculations are for a simpler situation than realized experimentally.
The calculations are for an ensemble of identical centers with identical optical pumping rates
whereas the experiment involves four orientations and variation in the optical pumping rates. The
consequence of these factors can be approximated by adding a square wave emission response to the
calculated emission response before comparing with experiment. The structured component of the
response ( the 'peak') will then represent a smaller fraction of each pulse. Another important
consideration is that photo-ionization has not been included. In the calculations the peak of the
second pulse is stronger than that of the first whereas it is the reverse in the experiment. The
difference is due to photo-ionization. Photo-ionization causes there to be a reduction in the number
of N-V centers during the first pulse (giving small alteration to the slope). The recovery is slow
and there is no recovery during the short dark period. Hence, the number of centers involved is
larger at the start of the first pulse than at the start of the second pulse. When allowing for these
factors and recognizing that the parameters have been determined from independent measurements the
correspondence between the calculated responses in Fig.9 and equivalent experimental traces in Fig.6
is very satisfactory.

\begin{figure}
\includegraphics[width=0.5\textwidth]{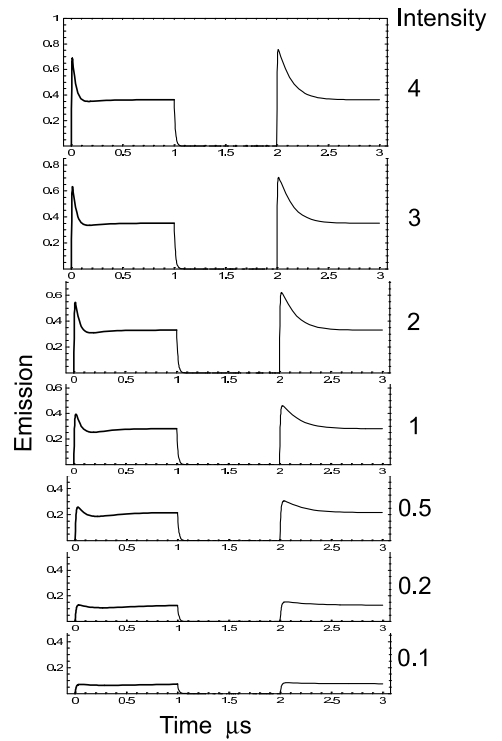}
\caption{(a) Emission predicted from the solution of the rate equations for a pair of excitation
pulses. The emission is shown for various excitation intensities k shown on the right in units of
1/$\tau$. The value of the parameters are given in the text and summarized in the caption of Fig. 8.}
\label{Fig9}
\end{figure}

\section{Magnetic field calculation}
A magnetic field other than along the trigonal axis causes mixing of the spin states \cite{Lou77} and
consequently with a randomly oriented field the populations are not associated with separate z, x and
y states. The effect can be approximated by retaining equal populations in the three spin projections
and the result of doing this is shown in Fig. 10. The upper trace gives the response in the absence
of a magnetic field and is the same as in the previous section with k = 1. For the lower trace the
populations in the three ground spin states are equalized. When this is done to approximate the
effect of a magnetic field, there is no spin polarization and the responses are the same for the two
pulses. With the field applied there continues to be optical excitation from the x, y state and more
efficient transfer into the singlet level. The result is that the equilibrium population in the
singlet level is higher and less centers contribute to the emission. The final emission is lower and
this is what is observed. It should be noted that equivalent effects can be obtained by applying
resonant microwave fields to maintain population in the x, y states. At high intensities the drop in
equilibrium emission level caused by the microwave field can be much larger \cite{Fedor} than the
14$\%$ obtained at low intensities. This is due to the change of the population stored in the singlet
level.

\begin{figure}
\includegraphics[width=0.5\textwidth]{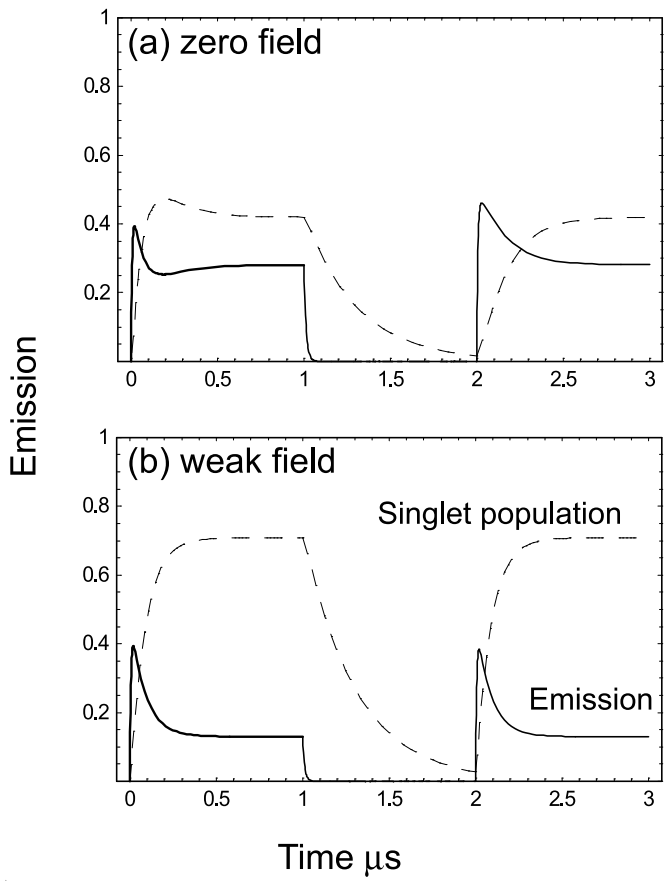}
\caption{Emission predicted from solutions of rate equations illustrating effect of an applied
magnetic field. The solid line in the upper trace shows the emission of system determined from
solution of rate equations for energy scheme as shown in Fig. 2. The dashed line indicates the
variation in population of singlet level. For the lower trace the three ground states are mixed,
effectively maintaining equal population in the three spin projections. The excitation rate is k = 1
in units of 1/$\tau$. The value of the parameters are given in the text and summarized in the caption
of Fig. 8.} \label{Fig10}
\end{figure}

In comparing the responses with and without an applied magnetic field there is a variation in the
magnitude of the peaks. The peak in the emission associated with the first pulse is similar with and
without field. However, the magnitudes associated with the second pulse are very different. As
discussed previously, the difference in peak heights between the first and second pulses is due to
photo-ionization varying the number of centers. A smaller second peak indicates that the magnetic
field has caused an increase in the photo-ionization. This can be attributed to the field increasing
the population in the excited states (excited triplet plus singlet) and the ionization being out of
these states. It is desirable to establish whether the ionization occurs through tunnelling out of
these states or is light induced. This requires further investigation.

\section{Comparison with other work}

There have been many publications referring to the singlet level in the N-V center and many of these
publications give information that is in conflict with the current model. The disparities are
discussed below.

In the present work it is shown that the singlet has a short lifetime of 0.3 $\mu$s. This contrasts
with the value of 0.275 s given when the intermediate $^{1}$A singlet was first proposed. Redman
et.al. \cite{Red91} deduced that a long lived electronic state accounted for a narrow 1.2 Hz
resonance observed in a near degenerate 4-wave mixing experiment. They recognized that the narrow
resonance could be associated with a long lived spin state but their estimates of the spin-lattice
relaxation times suggested otherwise. However, a spin-lattice relaxation time of 0.275 s is quite
realistic for N-V centers in diamond and we consider that this was the correct interpretation. In
this case their data can be explained without invoking a long lived singlet state and the information
would be consistent with that presented here.

The presence of a long lived singlet has also been adopted in accounting for the loss of emission
from single centers as temperature is lowered \cite{Dra99,Dra99a}. With excitation it was considered
that the singlet becomes populated but there is a thermally stimulated back transfer to the emitting
level such that at room temperature there is little loss of emission. The back transfer decreases
with lowering temperature and the center can remain in the singlet level for a considerable time
resulting in a drop of the average emission intensity. This decrease in intensity occurs for
zero-phonon line excitation. However, if the explanation is correct there will be an equivalent loss
of emission when the excitation is within the vibronic band. This is not the case. There is little
change in N-V emission intensity with a lowering of temperature (see Fig.3). The more likely
explanation for the loss of emission in the case of zero-phonon line excitation is spectral
hole-burning. There are a range of processes (change of spin state, re-orientation of center or
movement of charge in the neighborhood of the center) which can shift the absorption frequency and
cause a decrease of absorption for a laser held at fixed frequency. With a decrease in absorption
there will be an equivalent decrease in emission and the effect will become more pronounced as the
temperature is lowered due to narrowing of the homogeneous line width. The decrease in emission is,
therefore, attributed to this process and not with populating the singlet level. Clearly no
information about the energy of a singlet can be obtained from such experiments. \cite{Dra99a,Jel01}.

It is often assumed that there is no decay from the singlet to the ground state \cite{Kur00,Bro00}.
Should this be the case the spin polarization would have to arises through the back transfer and such
a process would be strongly temperature dependent. This is not what is observed. Spin polarization
occurs from liquid helium temperatures to room temperatures and in the original measurements of spin
polarization Loubser and van Wyk \cite{Lou77} have shown that the polarization is maintained to 500
K. Another more important issue is to question how spin polarization arises with the thermal back
transfer process. No details have been presented as to how the spin polarization occurs. This is in
contrast with the present model where, rather than back transfer, there is decay from the singlet
direct to the ground state. With the decay channels proposed one can readily account for the spin
polarization. It can be concluded that the N-V center can be understood without back transfer playing
a role.

It is noted that the analysis of the photon statistics associated with emission from single N-V
centers \cite{Dra99,Dra99a,Kur00,Bro00,Niz03} has assumed that either there is significant
singlet-triplet back transfer or a long lived singlet state (or both). Consequently the parameters
reported are inconsistent with the values obtained here. Such measurements need to be re-analyzed
using the present model and consideration given to contributions associated with photo-ionization.

Wrachtrup and co-workers at University of Stuttgart and at the National Academy of Sciences of
Belarus have presented an exciting range of single site measurements including demonstrating aspects
of quantum computing \cite{Wra01,Jel02,Niz03,Jel04b,Jel04a,Jel04,Pop04,Niz05,Wra05}. In the model
adopted to interpret their results they have spin as a good quantum number and spin-orbit interaction
is neglected. It is doubtful that spin-orbit can be totally neglected but it is small and there is a
correspondence in the energy level schemes and selection rules between their work and that given
here. There are also some similarities in the values of the parameters. For example, we have
determined that the rate constant for inter-system crossing from the excited x',y' states to the
singlet s has a value of k$_{x's}$ = k$_{y's}$ = 0.39 x 1/$\tau$. This is in good correspondence with
a value of 0.5 x 1/$\tau$  given by Nizovtsev et.al. \cite{Niz05}. They have also proposed that there
is much slower transfer from the $^{3}$E z' state to the singlet and we agree with this conclusion.
They have given a value of k$_{z's}$ = 2.5 x 10$^{-4}$ x 1/$\tau$. In our model it is considered to
be zero but a small value such as they have given does not change the behavior of the system. There
is, therefore, reasonable agreement when considering the populating of the singlet level. The
situation for transfer out of the singlet level is different and there is some confusion. There is no
agreement when the singlet is taken to have the long 0.275 s lifetime and when there is no decay from
this singlet to the ground state \cite{Niz05}. In this situation the spin polarization and transfer
out of the singlet has to be through thermal back transfer plus optically driven processes. However,
these processes are not consistent with the optically induced spin polarization being independent of
temperature and linear \cite{Hir92,Har04}. As there are no such processes in our model there can be
no comparison made with the parameters given. In other work \cite{Niz03a,Niz03,Jel04,Wra05} direct
transfer from the singlet to the ground state is indicated and mention made of a short singlet
lifetime \cite{Wra05}. No parameters are given to enable a comparison but clearly there is a
consistency with the model presented here.

Jelezko et al \cite{Jel04} have observed a single sharp zero-phonon line in the excitation spectrum
of single centers. This is a very significant observation as it is crucial for readout for many N-V
quantum information processing applications. Detection requires there to be a transition that cycles
without change of spin projection and their observation indicates that the z $\leftrightarrow$ z'
transition can cycle $\sim$100,000 x before the change occurs \cite{Wra05}. This is in contrast to
that obtained by our model, where with present parameters, the cycling of the z $\leftrightarrow$ z'
transition would be limited to $\sim$50 x before a change of spin state. In the model the cycling is
limited by the non-spin-conserving optical transitions and, as noted earlier, the strength of these
transitions vary with strain.  Strain varies the separation of S$_{z}$ and (S$_{x}$, S$_{y}$) spin
levels and consequently the degree of mixing via the non-axial spin-orbit interaction. This variation
has been investigated experimentally by Santouri et al \cite{San06}. By studying small regions of an
irradiated crystal they were able to obtain an inhomogeneous line width (15 GHz) orders of magnitude
narrower than obtained previously. Changing the spacial location gave spectra for different
magnitudes of strain and the $^{3}$E splitting was resolved. Furthermore, clear hyperfine structure
associated with the various optical transitions was obtained from two-laser hole burning
measurements. Previous two-laser hole burning data indicated that the S$_{z}$ spin state was lowest
in both the lower and upper component of the strain-split $^{3}$E state. This is confirmed in the
recent hole burning measurements and it is shown that this is the case with even small strain
splittings of $<$ 10 GHz. With such strain fields the order of the levels is, therefore, dominated by
"spin-spin type" terms rather than by spin-orbit. The effect of this interaction at zero strain in
Fig. 1(a) is to displace the S$_{x}$, S$_{y}$ states with respect to the S$_{z}$. This increases the
spin separations in the upper branch and reduces the separation in the lower branch (or vice versa)
and with increasing strain there will be a crossing of the spin levels in the lower energy branch
(between Fig 1(a) and 1(b)). Thus, in the lower branch the levels are closer and give larger mixing
of S$_{z}$ and S$_{x}$, S$_{y}$ spin states. Transitions to the lower branch are, therefore,
conducive to hole burning and electromagnetic induced transparency \cite{San06} as both require
non-spin conserving transitions. Alternatively transitions to the upper level are more favorable for
cyclic transitions and with low strain such transitions may account for the very cyclic transitions
observed by Jelezko et. al. \cite{Jel04}. An understanding of these processes and their variability
require knowledge of the magnitude of the interactions associated with the $^{3}$E excited state.
This has not been obtained in detail and remains an outstanding issue for a full understanding of the
electronic structure of the N-V center.

\section{Conclusion}

In this work group theoretical considerations have been used to account for the electronic states of
the N-V center, for the optical transitions and for the inter-system crossing. The account leads to a
seven level model where the dynamics are dominated by four rate constants. These four rate constants
are determined from independent experimental measurements and it is shown that with the values
obtained the model gives plausible correspondence with additional optical measurements of ensembles.
The comparison between theory and experiment is not fully quantitative but the agreement is
sufficient to give confidence in the appropriateness of the model. A good physical understanding of
the response of the center to optical excitation is obtained and the significance is that the model
provides a basis for the development of strategies to target the remaining outstanding issues
regarding the properties of the N-V center. The model also provides a sufficient understanding of the
dynamics to allow for satisfactory development of many N-V applications.

\section{Acknowledgements}
The above work has been supported by a DARPA QuIST grant through Texas Engineering Experimental
Station and by Australian Defence Science and Technology Organization. The authors thank Professors
Philip Hemmer (Texas A$\&$M University) and Fedor Jelezko (University of Stuttgart) for useful
discussions.

\end{document}